%% file: 00-main.tex
\documentclass[sigconf,dvipsnames]{acmart}
\usepackage[utf8]{inputenc}
\usepackage[T1]{fontenc}
\usepackage[inline]{enumitem}
\usepackage{diagbox}
\usepackage{booktabs}
\usepackage{amsmath}
\usepackage{amsfonts}
\usepackage{dashbox}
\usepackage{enumitem}
\usepackage{mathbbol}
\usepackage{mathtools}
\usepackage{hyperref}
\usepackage{url}
\usepackage{nicefrac}
\usepackage{microtype}
\usepackage{natbib}
\usepackage{multirow}
\usepackage{graphicx}
\usepackage{subcaption}
\usepackage{etaremune}
\usepackage{xcolor}
\usepackage{bbm}
\usepackage{algorithm}
\usepackage{algorithmic}
\usepackage{balance}

\newcommand{\ie}{\emph{i.e.}}
\newcommand{\eg}{\emph{e.g.}}
\newcommand{\wrt}{\emph{w.r.t. }}

\newcommand{\vs}{\emph{vs. }}

\makeatletter

\newcommand{\scriptshortto}[1][3pt]{{%
    \hbox{\rule[\scriptratio\dimexpr\fontdimen22\textfont2-.2pt\relax]
               {\scriptratio\dimexpr#1\relax}{\scriptratio\dimexpr.4pt\relax}}%
   \mkern-4mu\hbox{\let\f@size\sf@size\usefont{U}{lasy}{m}{n}\symbol{41}}}}

\makeatother

\AtBeginDocument{%
  \providecommand\BibTeX{{%
    \normalfont B\kern-0.5em{\scshape i\kern-0.25em b}\kern-0.8em\TeX}}}

\newcommand{\subtlesection}[1]{\smallskip\noindent\textbf{\emph{#1}.}}
\newcommand{\resq}[1]{\smallskip\noindent\textbf{#1}}

\setcopyright{acmcopyright}
\copyrightyear{2018}
\acmYear{2018}
\acmDOI{10.1145/1122445.1122456}

\acmConference[Woodstock '18]{Woodstock '18: ACM Symposium on Neural
  Gaze Detection}{June 03--05, 2018}{Woodstock, NY}
\acmBooktitle{Woodstock '18: ACM Symposium on Neural Gaze Detection,
  June 03--05, 2018, Woodstock, NY}
\acmPrice{15.00}
\acmISBN{978-1-4503-XXXX-X/18/06}

\begin{document}

\setlength{\abovedisplayskip}{5pt}
\setlength{\belowdisplayskip}{5pt}
\setlength{\abovedisplayshortskip}{5pt}
\setlength{\belowdisplayshortskip}{5pt}
\addtolength{\parskip}{-0.5mm}


\title{Improving Transformer-Kernel Ranking Model Using Conformer and Query Term Independence}

\author{Bhaskar Mitra}
\affiliation{%
  \institution{Microsoft}
}
\email{bmitra@microsoft.com}

\author{Sebastian Hofst\"{a}tter}
\affiliation{%
  \institution{TU Wien}
}
\email{s.hofstaetter@tuwien.ac.at}

\author{Hamed Zamani}
\affiliation{%
  \institution{University of Massachusetts Amherst}
}
\email{zamani@cs.umass.edu}

\author{Nick Craswell}
\affiliation{%
  \institution{Microsoft}
}
\email{nickcr@microsoft.com}

\input{01-abstract}
\maketitle

\input{02-intro}
\input{03-related}
\input{04-model}
\input{05-experiment}
\input{06-result}

\input{07-conclusion}

\bibliographystyle{ACM-Reference-Format}
\bibliography{bibtex}

\end{document}

%% file: 01-abstract.tex
\begin{abstract}
The Transformer-Kernel (TK) model has demonstrated strong reranking performance on the TREC Deep Learning benchmark---and can be considered to be an efficient (but slightly less effective) alternative to other Transformer-based architectures that employ
\begin{enumerate*}[label=(\roman*)]
    \item large-scale pretraining (high training cost),
    \item joint encoding of query and document (high inference cost), and
    \item larger number of Transformer layers (both high training and high inference costs).
\end{enumerate*}
Since, a variant of the TK model---called TKL---has been developed that incorporates local self-attention to efficiently process longer input sequences in the context of document ranking.
In this work, we propose a novel Conformer layer as an alternative approach to scale TK to longer input sequences.
Furthermore, we incorporate query term independence and explicit term matching to extend the model to the full retrieval setting.
We benchmark our models under the strictly blind evaluation setting of the TREC 2020 Deep Learning track and find that our proposed architecture changes lead to improved retrieval quality over TKL.
Our best model also outperforms all non-neural runs (``trad'') and two-thirds of the pretrained Transformer-based runs (``nnlm'') on NDCG@10.
\end{abstract}

%% file: 02-intro.tex
\section{Introduction}
\label{sec:intro}

In the inaugural year of the TREC Deep Learning track~\citep{trec2019overview}, ranking models using Transformers~\citep{vaswani2017attention} demonstrated substantial improvements over traditional information retrieval (IR) methods.
Several of these approaches---\eg, \citep{yilmaz2019h2oloo, yan2019idst}---employ BERT~\citep{devlin2018bert}, with large-scale pretraining, as their core architecture.
Diverging from this trend, \citet{Hofstaetter2020_ecai} propose the Transformer-Kernel (TK) model with few key distinctions:
\begin{enumerate*}[label=(\roman*)]
    \item TK uses a shallower model with only two Transformer layers,
    \item there are no computation-intensive pretraining, and
    \item TK independently encodes the query and document allowing for offline precomputations for faster response times.
\end{enumerate*}
Consequently, TK achieves competitive performance at a fraction of the training and inference cost of its BERT-based peers.

Notwithstanding these efficiency gains, the TK model shares two critical drawbacks with other Transformer-based models.
Firstly, the memory complexity of the self-attention layers is quadratic $\mathcal{O}(n^2)$ with respect to the length $n$ of the input sequence.
This restricts the number of document terms we can inspect under fixed GPU memory budget.
A trivial workaround involves inspecting only the first $k$ terms of the document.
This approach can negatively impact retrieval quality and has been shown to under-retrieve longer documents~\citep{hofstatter2020improving}.
Secondly, in any real IR system, it is impractical to exhaustively evaluate every document in the collection for every query---and therefore these systems typically enforce some sparsity property to drastically narrow down the set of candidates.
TK employs a nonlinear matching function over query-document pairs which makes it difficult to enforce such sparsity before model inference.
This restricts TK's scope of application to late stage reranking of smaller candidate sets as identified by simpler retrieval models.
So, in this work, we extend TK in the following ways:
\begin{enumerate}
    \item To scale to long text, we replace the Transformer layers with novel Conformer layers whose memory complexity is $\mathcal{O}(n \times d_\text{key})$, instead of $\mathcal{O}(n^2)$,
    \item To enable fast retrieval with TK, we incorporate query term independence (QTI)~\citep{mitra2019incorporating}, and finally,
    \item we complement TK's latent matching with lexical term matching as suggested previously by \citet{mitra2016desm, mitra2017learning}, which is known to be effective for full retrieval~\citep{mitra2016desm, kuzi2020leveraging, gao2020complementing, wrzalik2020cort}.
\end{enumerate}

We study the impact of aforementioned changes under the strictly-blind evaluation setting of the TREC 2020 Deep Learning track.

%% file: 03-related.tex
\section{Related work}
\label{sec:related}

\subtlesection{Scaling self-attention to long text}
\label{sec:related-long}
The self-attention layer, as proposed by \citet{vaswani2017attention}, can be described as follows:
\begin{align}
    \text{Self-Attention}(Q, K, V) &= \Phi(\frac{QK^\intercal}{\sqrt{d_k}}) \cdot V
\end{align}
Where, $Q \in \mathbb{R}^{n \times d_\text{key}}$, $K \in \mathbb{R}^{n \times d_\text{key}}$, and $V \in \mathbb{R}^{n \times d_\text{value}}$ are the query, key, and value matrices---and $d_\text{key}$ and $d_\text{value}$ are the dimensions of the key and value embeddings, respectively.
Here, $n$ is the length of the input sequence and $\Phi$ denotes a softmax along the last tensor dimension.
The quadratic $\mathcal{O}(n^2)$ memory complexity of self-attention is a direct consequence of the component $QK^\intercal$ that produces a  $n \times n$ matrix.
Recently, several approaches have been proposed to mitigate this quadratic complexity that broadly fall under:
\begin{enumerate*}[label=(\roman*)]
    \item Restricting self-attention to smaller local windows over the input~\citep{parmar2018image, dai2019transformer, yang2019xlnet, sukhbaatar2019adaptive}, or
    \item operating under the assumption that the attention matrix is low rank $r$~\citep{kitaev2019reformer, roy2020efficient, tay2020sparse, wang2020linformer} and hence finding alternatives to explicitly computing the $QK^\intercal$ matrix, or
    \item hybrid approaches~\citep{child2019sparse, beltagy2020longformer, Wu2020Lite}.
\end{enumerate*}
In IR, recently \citet{hofstatter2020improving} extended TK to longer text using local self-attention.
Other more general approaches to reducing the memory footprint, such as model parallelization~\citep{shoeybi2019megatron} and gradient checkpointing~\citep{beltagy2020longformer} have also been explored.

\subtlesection{Full retrieval with deep models}
\label{sec:related-full}
Efficient retrieval using deep models is an important challenge in IR~\citep{mitra2018introduction, mitra2021neural}.
One approach involves the dual encoder architecture where the query and document are encoded independently, and efficient retrieval is achieved by approximate nearest-neighbour search~\citep{lee2019latent, chang2020pre, karpukhin2020dense, ahmad2019reqa, khattab2020colbert} or by employing inverted-index over latent representations~\citep{zamani2018neural2}.
Precise matching of terms or concepts may be difficult using query-independent latent document representations~\citep{luan2020sparse}, and therefore these models are often combined with explicit term matching~\citep{nalisnick2016improving, mitra2017learning}.

An alternative approach assumes QTI in the design of the neural ranking model~\citep{mitra2019incorporating}.
In these models, the estimated relevance score $S_{q,d} = \sum_{t \in q}{s_{t,d}}$ is the sum of the document scores \wrt individual query terms.
Readers should note that QTI is already baked into several classical IR models, like BM25~\citep{robertson2009probabilistic}.
Relevance models with QTI can be used to offline precompute all term-document scores, and subsequently efficient search is performed using inverted-index.
Several recent neural IR models~\citep{mitra2019incorporating, dai2019evaluation, dai2019deeper, mackenzie2020efficiency, daicontext, macavaney2020expansion} that incorporate QTI have obtained promising results under the full retrieval setting.
Document expansion based methods~\citep{nogueira2019document, nogueira2019doc2query}, using large neural language models, can also be classified as part of this approach, assuming the subsequent retrieval step employs a traditional QTI model like BM25.
In all these cases, the focus of the deep model is to estimate the relevance of the document \wrt individual terms in the vocabulary that can be precomputed during indexing.
Another approach may involve neural query reformulation~\citep{nogueira2017task, van2017reply, ma2020zero}, although these methods typically underperform compared to the methods considered here.


%% file: 04-model.tex
\section{Conformer-Kernel with QTI}
\label{sec:model}

\begin{figure*}
\center
\begin{subfigure}{\columnwidth}
    \includegraphics[width=\textwidth]{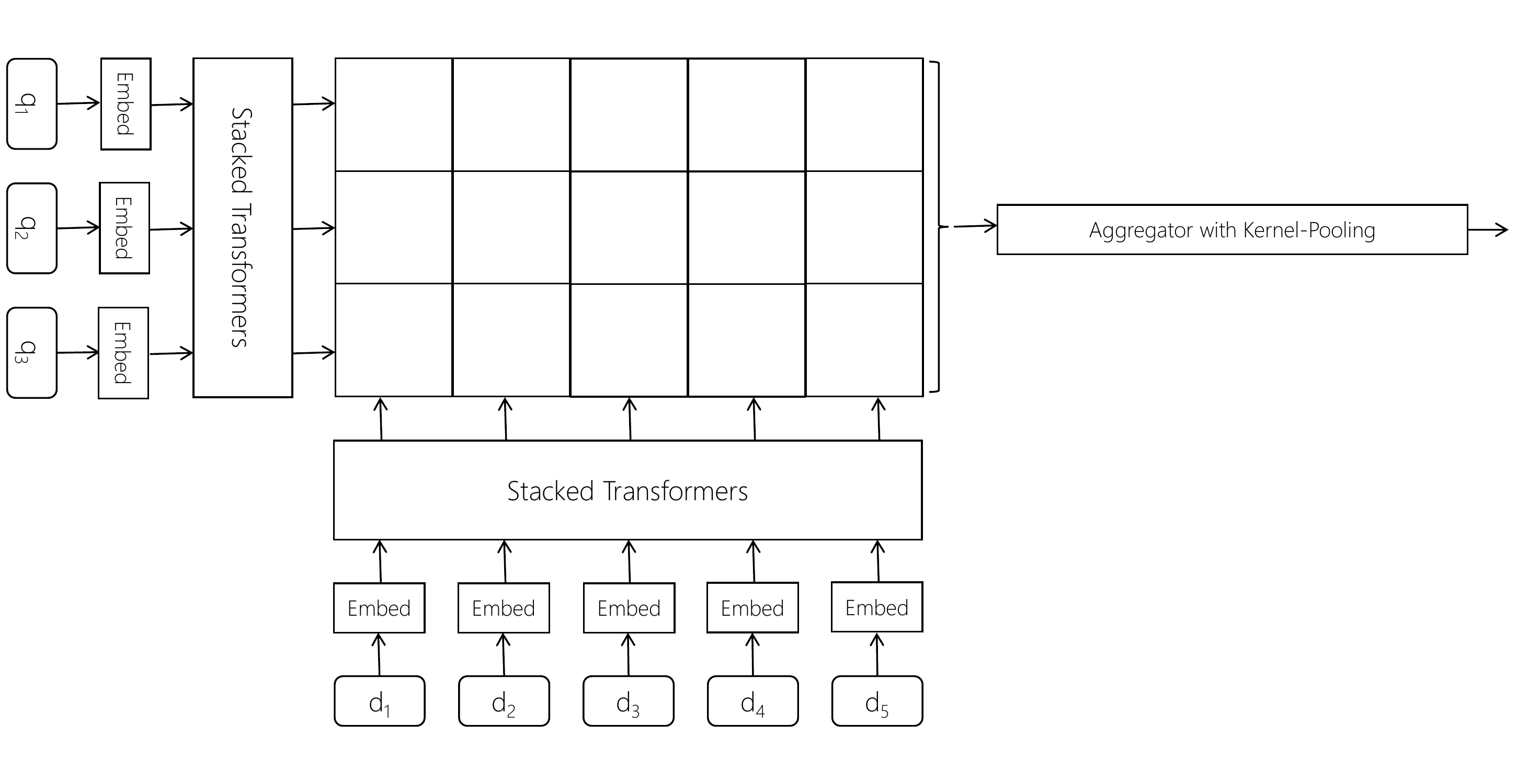}
    \caption{Transformer-Kernel (TK)}
    \label{fig:model-tk}
\end{subfigure}
\hfill
\begin{subfigure}{\columnwidth}
    \includegraphics[width=\textwidth]{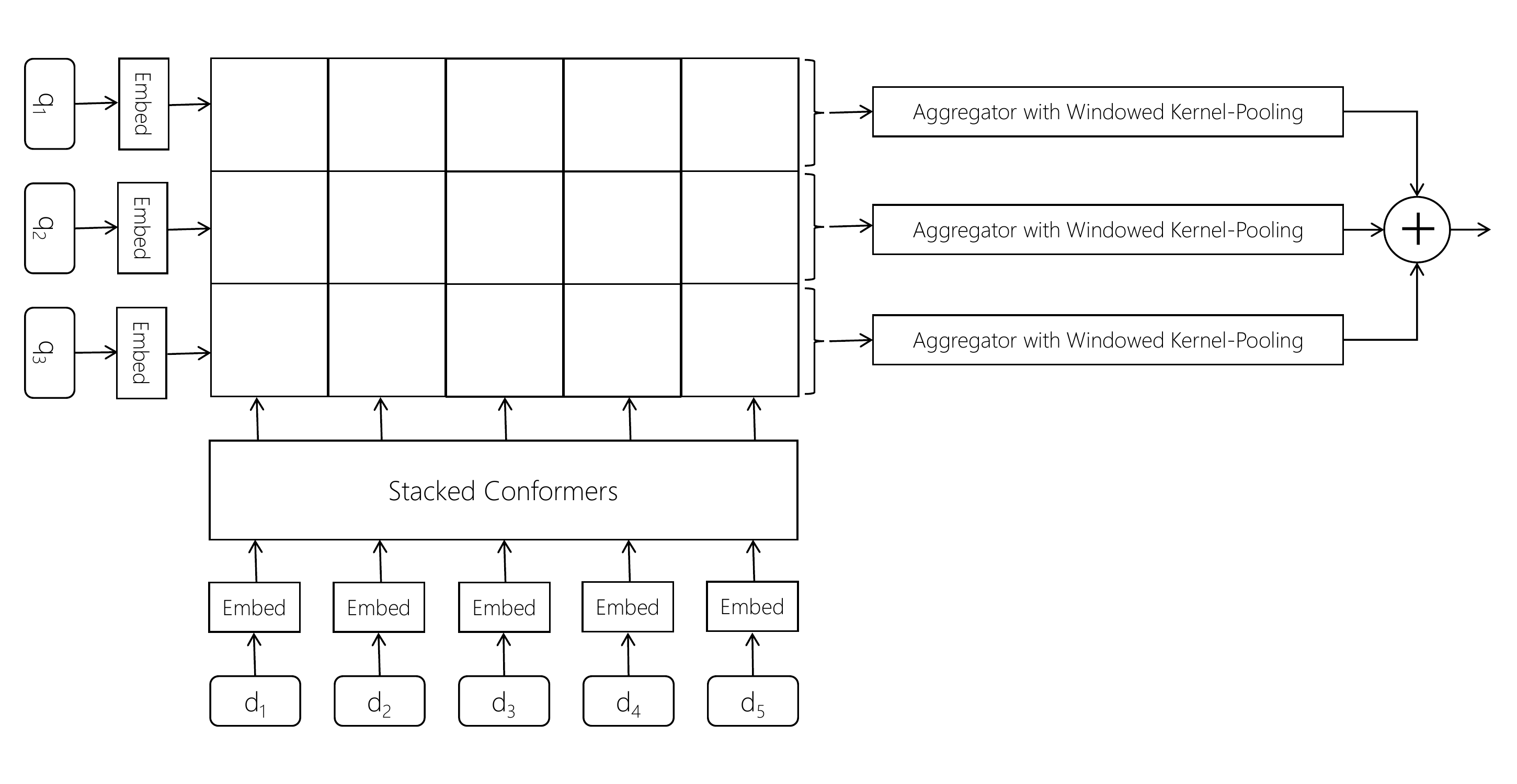}
    \caption{NDRM1 variant of Conformer-Kernel (CK) with QTI}
    \label{fig:model-ck}
\end{subfigure}
\caption{A comparison of the TK and the proposed CK-with-QTI architectures.
In addition to replacing the Transformer layers with Conformers, the latter also simplifies the query encoding to non-contextualized term embedding lookup and incorporates a windowed Kernel-Pooling based aggregation that is employed independently per query term.
}
\label{fig:model}
\end{figure*}

%
%
%
%

\subtlesection{Conformer}
\label{sec:model-conformer}
The quadratic memory complexity of self-attention layers \wrt the input length is a direct result of explicitly computing the attention matrix $QK^\intercal \in \mathbb{R}^{n \times n}$.
In this work, we propose a new separable self-attention layer that avoids instantiating the full term-term attention matrix.
\begin{align}
    \text{Separable-Self-Attention}(Q, K, V) &= \Phi(Q) \cdot A 
\end{align}
Where, $A = \Phi(K^\intercal) \cdot V$.
As previously, $\Phi$ denotes softmax along the last dimension of the input tensor.
Note that, however, in this separable self-attention mechanism, the softmax operation is employed twice:
\begin{enumerate*}[label=(\roman*)]
    \item $\Phi(Q)$ computes the softmax along the $d_\text{key}$ dimension, and
    \item $\Phi(K^\intercal)$ computes the softmax along the $n$ dimension.
\end{enumerate*}
By computing $A \in \mathbb{R}^{d_\text{key} \times d_\text{value}}$ first, we avoid explicitly computing the full term-term attention matrix.
The memory complexity of the separable self-attention layer is $\mathcal{O}(n \times d_\text{key})$, which is a significant improvement when $d_\text{key} \ll n$.
We modify the standard Transformer block as follows:
\begin{enumerate*}[label=(\roman*)]
    \item We replace the standard self-attention layer with the more memory efficient separable self-attention layer, and
    \item we apply grouped convolution before the separable self-attention layers to better capture the local context based on the window of neighbouring terms.
\end{enumerate*}
We refer to this combination of grouped \underline{con}volution and Trans\underline{former} with separable self-attention as a Conformer.
We incorporate Conformers into TK as a direct replacement for the Transformer layers and name the new architecture as a Conformer-Kernel (CK) model.
In relation to handling long input sequences, we also replace the standard Kernel-Pooling with windowed Kernel-Pooling~\citep{hofstatter2020improving} in our proposed architecture.

\subtlesection{Query term independence}
\label{sec:model-qti}
To incorporate QTI into CK, we make two simple modifications.
Firstly, we simplify the query encoder by getting rid of the Transformer layers and only considering the non-contextualized embeddings for the query terms.
Secondly, instead of applying the aggregation function over the full interaction matrix, we apply it to each row individually, which corresponds to individual query terms.
The scalar outputs from the aggregation function are linearly combined to produce the final query-document score.
Fig~\ref{fig:model-ck} shows the proposed CK-QTI architecture.

\subtlesection{Explicit term matching}
\label{sec:model-duet}
We adopt the Duet~\citep{mitra2017learning, nanni2017benchmark, mitra2019updated, mitra2019duet} framework wherein the term-document score is a linear combination of outputs from a latent and and an explicit matching models.
\begin{align}
    s_{t,d} = w_1 \cdot \text{BN}(s_{t,d}^\text{(latent)}) + w_2 \cdot \text{BN}(s_{t,d}^\text{(explicit)}) + b
\end{align}
Where, $\{w_1, w_2, b\}$ are learnable parameters and $\text{BN}(x) = (x - \mathbb{E}[x])/(\sqrt{\text{Var}[x]})$ denotes the BatchNorm operation~\citep{ioffe2015batch}.
We employ CK and define a new lexical matching function modeled on BM25 to compute $s_{t,d}^\text{(latent)}$ and $s_{t,d}^\text{(explicit)}$, respectively.
\begin{align}
    s_{t,d}^\text{(explicit)} &= \text{IDF}_{t} \cdot \frac{\text{BS}(\text{TF}_{t,d})}{\text{BS}(\text{TF}_{t,d}) + \text{ReLU}(w_\text{dlen} \cdot \text{BS}(|d|) + b_\text{dlen}) + \epsilon}
    \label{eqn:model-explicit}
\end{align}
Where, $\text{IDF}_{t}$, $\text{TF}_{t,d}$, and $|d|$ denote the inverse-document frequency of the term $t$, the term-frequency of $t$ in document $d$, and the length of the document, respectively.
The $w_\text{dlen}$ and $b_\text{dlen}$ are the only two leanrable parameters of this submodel and $\epsilon$ is a small constant added to prevent a divide-by-zero error.
The BatchScale (BS) operation is defined as $\text{BS}(x) = x/(\mathbb{E}[x] + \epsilon)$.

%% file: 05-experiment.tex
\section{Experiment design}
\label{sec:experiment}

\subtlesection{TREC 2020 Deep Learning Track}
\label{sec:experiment-data}
We evaluate CK under the strictly-blind TREC benchmarking setting\footnote{
We exclude group name and run IDs here to anonymize for the blind-review process.}
by participating in the 2020 edition of the Deep Learning track~\citep{trec2020overview}, which:
\begin{enumerate*}[label=(\alph*)]
    \item provides stronger protection against overfitting that may result from the experimenter running multiple evaluations against the test set, and
    \item is fairer to dramatically new approaches that may surface additional relevant documents not covered by pre-collected labels~\citep{yilmaz2020reliability}.
\end{enumerate*}
The 2020 track~\citep{trec2020overview} uses the same training data as the previous year~\citep{trec2019overview} originally derived from the MS MARCO dataset~\citep{bajaj2016ms}.
However, the track provides a new blind test set for the second year.
We only focus on the document ranking task and point the reader to~\citep{trec2020overview} for further benchmarking details.
We report NDCG@10~\citep{JK2002}, NCG@100~\citep{rosset2018optimizing}, AP~\citep{zhu2004recall}, and RR~\citep{craswell2009mean} against this blind set.


\subtlesection{Model variants}
\label{sec:experiment-model}
We compare several variants of our model.
The \emph{NDRM1} variant incorporates Conformer layers and QTI into TK~\citep{Hofstaetter2020_ecai}.
Figure~\ref{fig:model} visualizes the NDRM1 architecture.
The \emph{NDRM2} model is a simple QTI-compliant explicit-term-matching model as described by Equation~\ref{eqn:model-explicit}.
A linear combination of NDRM1 and NDRM2 gives us the NDRM3 model.
Because of the limit on the number of run submission to TREC, we only evaluate NDRM1 and NDRM3, although we confirm on the TREC 2019 test set that NDRM2 is competitive with a well-tuned BM25 baseline.
The TREC 2020 Deep Learning track provided participants with a click log dataset called ORCAS~\citep{craswell2020orcas}.
We use clicked queries in the ORCAS data~\citep{craswell2020orcas} as additional meta description for corresponding documents to complement the intrinsic document content (URL, title, and body).
Unlike previous work~\citep{zamani2018neural2} on fielded document representations, we simply concatenate the different fields.
We test each variant under both the rerank and the fullrank settings.

\subtlesection{Model training}
\label{sec:experiment-training}
We consider the first $20$ terms for every query and the first $4000$ terms for every document.
We pretrain the word embeddings using the word2vec~\citep{mikolov2013distributed} implementation in FastText~\citep{joulin2016bag}.
We use a concatenation of the IN and OUT embeddings~\citep{nalisnick2016improving, mitra2016desm} from word2vec to initialize the embedding layer parameters.
The document encoder uses 2 Conformer layers and we set all hidden layer sizes to $256$.
We set the window size for the grouped convolution layers to $31$ and the number of groups to $32$.
Correspondingly, we also set the number of attention heads to $32$.
We set the number of kernels $k$ to $10$.
For windowed Kernel-Pooling, we set the window size to $300$ and the stride to $100$.
Finally, we set the dropout rate to $0.2$.
For further details, please refer to the publicly released model implementation in PyTorch.\footnote{
We will add a link to the repo here after completion of the blind-reviewing process.
}
All models are trained on four Tesla P100 GPUs, with 16 GB memory each, using data parallelism.

We train the model using the RankNet objective~\citep{burges2005learning}.
For every positively labeled query-document pair in the training data, we randomly sample one negative document from the provided top $100$ candidates corresponding to the query and two negative documents from the full collection.
In addition to making pairs between the positively labeled document and the three negative documents, we also create pairs between the negative document sampled from the top $100$ candidates and those sampled from the full collection, treating the former as more relevant.
This can be interpreted as incorporating a form of weak supervision~\citep{dehghani2017neural} as the candidates were previously generated using a traditional IR function.

%% file: 06-result.tex
\section{Results}
\label{sec:result}

\begin{table}
    \small
    \centering
    \caption{Official TREC 2020 results.
    All metrics are computed at rank $100$, except for NDCG which is computed at rank $10$.
    Best and median runs are selected based on NDCG@10.}
    \begin{tabular}{llllll}
    \hline
    \hline
        \textbf{Run description} & \textbf{Subtask} & \textbf{NDCG} & \textbf{NCG} & \textbf{AP} & \textbf{RR} \\
        \hline
        \multicolumn{6}{l}{\textbf{Other TREC runs for comparison}} \\
        Best ``trad'' run & fullrank & $0.5629$ & $0.6299$ & $0.3829$ & $0.9195$ \\
        Best TKL run & rerank & $0.5852$ & $0.6283$ & $0.3810$ & $0.9296$ \\
        Median ``nnlm'' run & fullrank & $0.5907$ & $0.6669$ & $0.4259$ & $0.8916$ \\
        Best ``nnlm'' run & fullrank & $0.6934$ & $0.7718$ & $0.5422$ & $0.9476$ \\
        \hline
        \multicolumn{6}{l}{\textbf{Our models}} \\
        NDRM1 & fullrank & $0.5991$  & $0.6280$ & $0.3858$ & $0.9333$ \\
        NDRM1 & rerank & $0.6161$ & $0.6283$ & $0.4150$ & $0.9333$ \\
        NDRM3 & rerank & $0.6162$ & $0.6283$ & $0.4122$ & $0.9333$ \\
        NDRM3 & fullrank & $0.6162$ & $0.6626$ & $0.4069$ & $0.9333$ \\
        NDRM3 + ORCAS & rerank & $0.6217$ & $0.6283$ & $0.4194$ & $0.9241$ \\
        NDRM3 + ORCAS & fullrank & $0.6249$ & $0.6764$ & $0.4280$ & $0.9444$ \\
        \hline
        \hline
    \end{tabular}
    \label{tbl:results}
\end{table}

\resq{RQ1. Does CK-QTI improve reranking quality over TKL?}
According to the taxonomy proposed by \citet{trec2019overview}, CK-QTI and TKL runs are the only ``nn'' runs---\ie, neural models that do not use pretrained transformers---submitted to TREC 2020 Deep Learning track.
TKL has previously been shown to outperform TK~\citep{hofstatter2020improving}, and we confirmed with the submitting group that they considered these as well-tuned TKL runs.
We also confirm that the related hyperparameters are comparable between the TKL runs and ours.
Table~\ref{tbl:results} shows that in the same rerank setting, both NDRM1 and NDRM3 improve NDCG@10 over the best TKL run by $5.3\%$.
The improvement from NDRM1 over TKL is statistically significant according to student's t-test ($p < 0.05$).
However, similarly large improvement from NDRM3 over TKL is not stat. sig. likely due to small test set size.
Even if we consider TK and CK to be comparable in results quality, the key motivation behind Conformers is their reduced GPU memory usage which we discuss next.


\begin{figure}
\center
\includegraphics[width=\columnwidth]{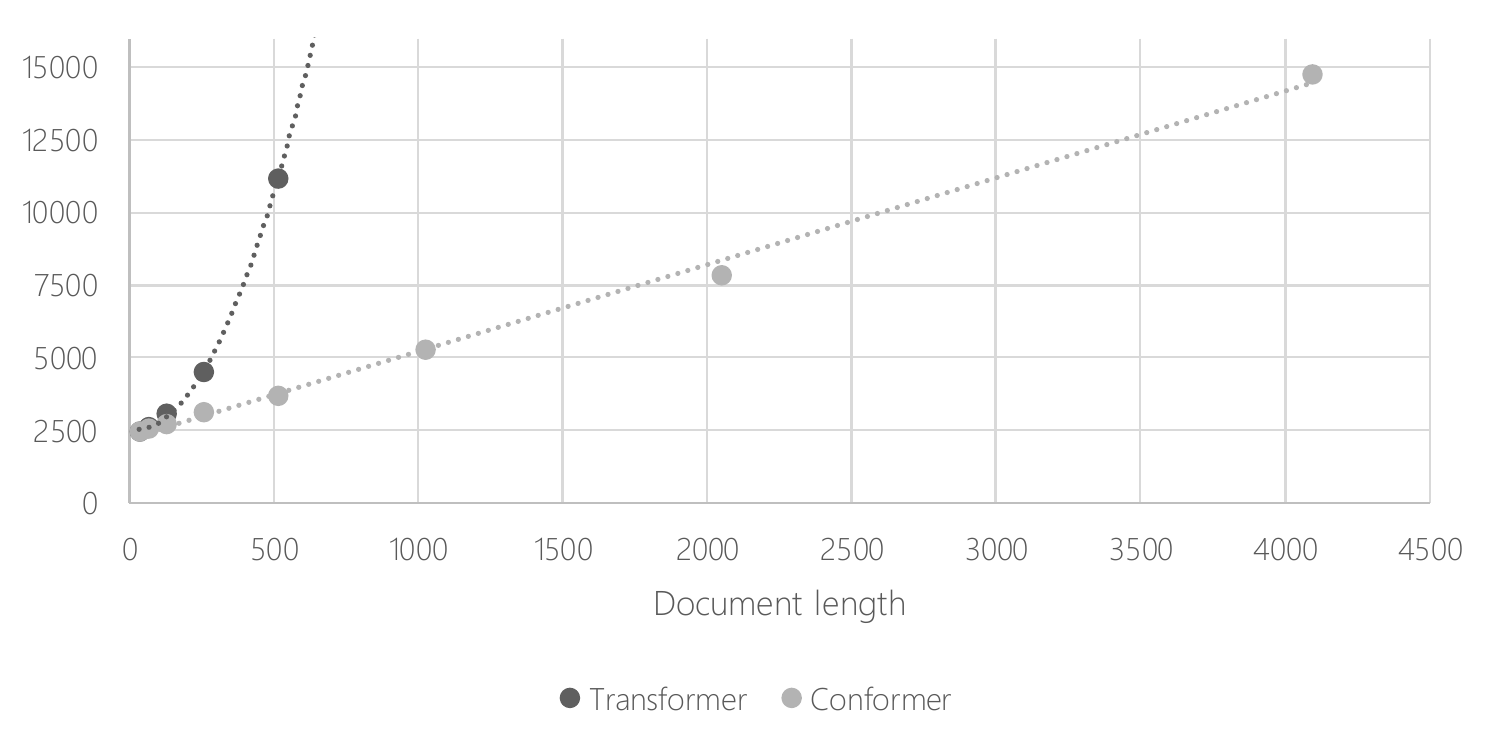}
\caption{Comparison of peak GPU Memory Usage in MB, across all four GPUs, when employing Transformers \vs Conformers.
}
\label{fig:memory}
\end{figure}

\resq{RQ2. Does CK-QTI improve train-time GPU memory requirement over TKL?}
To demonstrate how the GPU memory consumption scales with respect to input sequence length, we plot the peak memory, across all four GPUs, for our proposed architecture using Transformer and Conformer layers, respectively, keeping all other hyperparameters and architecture choices fixed.
Fig~\ref{fig:memory} shows the GPU memory requirement grows linearly with increasing sequence length for the Conformer, while quadratically when Transformers are employed.
This is a significant improvement in GPU memory requirement over TK for longer text that could be further operationalized to improve training time convergence using larger batches or to incorporate longer input representations of documents.

\begin{figure}
  \center
  \begin{subfigure}{.9\columnwidth}
    \includegraphics[width=\textwidth]{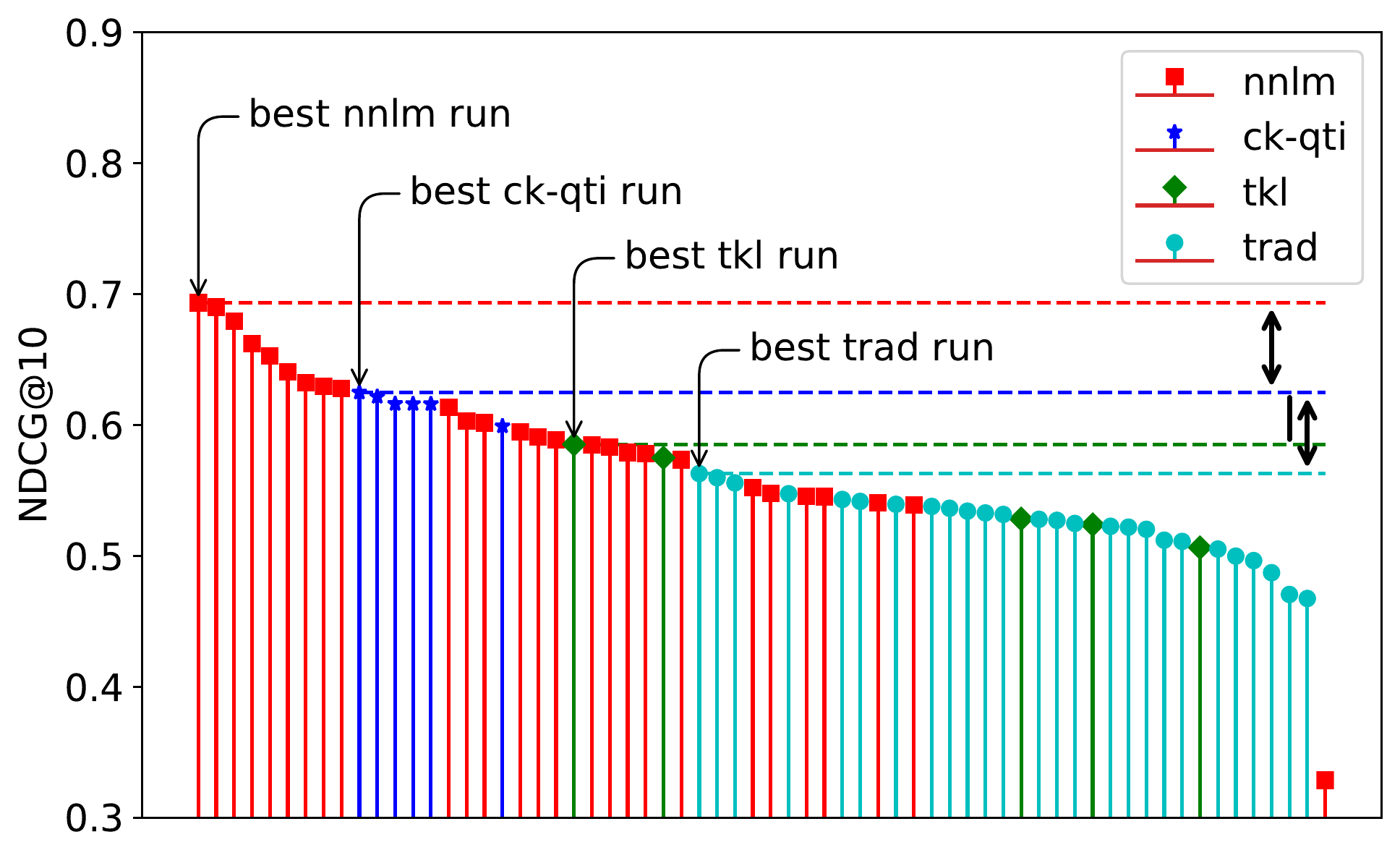}
    \caption{NDCG@10}
    \label{fig:competitive-ndcg}
  \end{subfigure}
  \hfill
  \begin{subfigure}{.9\columnwidth}
    \includegraphics[width=\textwidth]{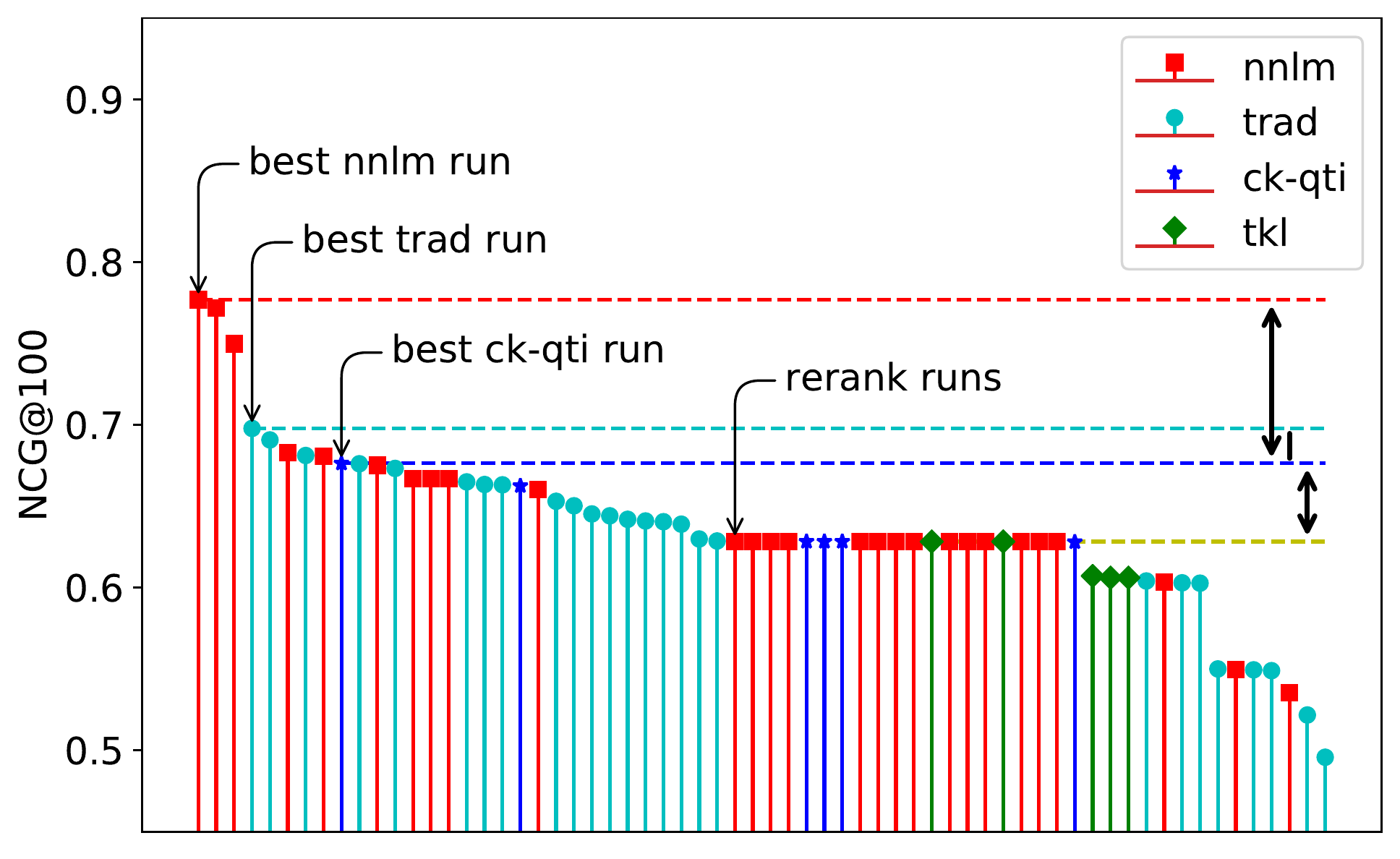}
    \caption{NCG@100}
    \label{fig:competitive-ncg}
  \end{subfigure}
  \addtolength{\belowcaptionskip}{-12pt}
  \caption{Comparing CK-QTI runs with runs submitted by other groups. 
  The runs in each plot are sorted independently based on the corresponding metric.}
  \label{fig:competitive}
\end{figure}

\resq{RQ3. How does CK-QTI perform in the full retrieval setting?}
To enable retrieval from the full collection, we incorporate two changes in TK: QTI and explicit term matching.
QTI allows for precomputation of term-document scores and consequently fast retrieval using inverted-index data structures.
The explicit term matching is expected to help with result quality under the full retrieval setting.
In Table~\ref{tbl:results}, we find that the NDRM3 variant---that incorporates explicit term matching---does indeed achieve $2.9\%$ better NDCG@10 compared to the NDRM1 variant and $5.5\%$ improvement in both AP and NCG@100.
In contrast, both models achieve similar performance under the rerank setting.
The candidate documents for reranking were generated by a first-stage BM25 ranker and hence explicit term matching signal is already part of this retrieval pipeline which may explain why we find no benefit from explicit term matching in reranking.
These observations are supported by \citet{kuzi2020leveraging}, who find that exact term matching are important for the fullrank setting.
Also, NDRM1, in the absence of explicit term matching, achieves a lower NDCG@10 under the fullrank setting compared to the rerank setting.
However, when explicit term matching is incorporated (\ie, NDRM3), the metrics are comparable under both settings.
Interestingly, when we include the ORCAS data in the document representation, we see improvements under the fullrank setting compared to reranking across all metrics: $2.2\%$ for RR, $2.1\%$ for AP, and $0.5\%$ for NDCG@10.
We confirm that the NDCG@10 improvement from fullrank over rerank setting under the NDRM3 + ORCAS configuration is stat. sig. based on a student's t-test ($p < 0.05$).
Based on qualitative inspection of the queries, we find that exact term matching may be important for queries containing named entities---\eg, ``who is \emph{aziz hashim}'' and ``why is \emph{pete rose} banned from hall of fame''---where it is necessary to ensure that the retrieved documents are about the correct entity.
Finally, with respect to the full retrieval setting, we note that NDRM3 with ORCAS improves NCG@100 by $7.7\%$ over the provided candidates for the reranking setting, which puts it among the 10 top performing runs according to NCG@100 as seen in Fig~\ref{fig:competitive}.

\resq{RQ4. How does CK-QTI compare to ``trad'' and ``nnlm'' runs?}
In adhoc retrieval, a common strategy involves sequentially cascading multiple rank-and-prune stages~\citep{matveeva2006high, wang2011cascade, chen2017efficient, gallagher2019joint, nogueira2019multi} for better effectiveness-efficiency trade-offs.
The multiple stages can improve result quality at additional computation costs.
However, in our experiments under the full retrieval setting, we employ CK-QTI as a single stage retriever.
Despite of this straightforward and efficient setup, we find that all three runs NDRM1, NDRM3, and NDRM3 + ORCAS achieve better NDCG@10 compared to the best non-neural (\ie, ``trad'') run.
The improvements from NDRM3, both with and without the ORCAS-based document representation, is stat. sig. compared to the best ``trad'' run based on student's t-test ($p < 0.05$). 
Additionally, NDRM3, with and without ORCAS, outperforms two-thirds of the ``nnlm'' runs that employ costly pretraining of Transformers.
The ``nnlm'' runs that outperform CK-QTI not only employ cascades of multiple rank-and-prune stages but sometimes multiple of those stages employ costly models like BERT.
In contrast, CK-QTI retrieves from the full collection in one-shot and its performance can be likely improved by additional reranking stages.




%% file: 07-conclusion.tex
\section{Conclusion}
\label{sec:conclusion}
We update TK by
\begin{enumerate*}[label=(\roman*)]
    \item replacing Transformers with Conformers, and incorporating
    \item QTI and
    \item explicit term matching.
\end{enumerate*}
Conformers scale better to longer inputs and show both relevance and GPU memory improvements.
Incorporating QTI and explicit term matching adapts the model to fullrank setting. In spite of being a single-stage retriever, CK-QTI outperforms all traditional methods and two-thirds of pretrained Transformer models.
We believe that CK, like its predecessor TK, represents an alternative to BERT-based ranking models at lower training and run-time inference cost.